\documentclass[aps,prb,twocolumn,superscriptaddress,showpacs,floatfix]{revtex4}
\usepackage{graphicx}
\begin{document}
\title{Chiral symmetry analysis and rigid rotational invariance
for the lattice dynamics of single-wall carbon nanotubes}
\author{Jin-Wu Jiang}
\affiliation{Institute of Theoretical Physics, Chinese Academy of
Sciences, Beijing 100080, China }
\author{Hui Tang}
\affiliation{Institute of Theoretical Physics, Chinese Academy of
Sciences, Beijing 100080, China }
\author{Bing-Shen Wang}
\affiliation{National Laboratory of Semiconductor Superlattice and
Microstructure
\\ and
Institute of Semiconductor, Chinese Academy of Sciences, Beijing
100083, China\\}
\author{Zhao-Bin Su}
\affiliation{Institute of Theoretical Physics, Chinese Academy of
Sciences, Beijing 100080, China }
\affiliation{Center for Advanced
Study, Tsinghua University, Beijing 100084, China}
\begin{abstract}
       In this paper,
       we provide a detailed expression of the vibrational potential for the
       lattice dynamics of the single-wall carbon nanotubes (SWCNT)
       satisfying the requirements of the exact rigid translational
       as well as rotational symmetries, which is a nontrivial
       generalization of the valence force model for the planar
       graphene sheet. With the model, the low frequency behavior of the
       dispersion of the acoustic modes as well as the flexure mode
       can be precisely calculated. Based upon
       a comprehensive chiral symmetry analysis, the calculated
       mode frequencies (including all the Raman and infrared active modes),
       velocities of acoustic modes and the polarization vectors are
       systematically fitted in terms of the chiral angle and
       radius, where the restrictions of various symmetry
       operations of the SWCNT are fulfilled.
\end{abstract}

 \maketitle
\section{INTRODUCTION}
Since the discovery\cite{Iijima} of carbon nanotubes, it
stimulated extensive studies on the lattice dynamics of the
single-wall carbon nanotubes
(SWCNT).\cite{Dresselhaus2,Dresselhaus6,Dresselhaus1,Dresselhaus3,Dresselhaus5}
The SWCNT can be viewed as rolling up a two-dimensional graphitic
lattice into a seamless cylinder.\cite{White} Among various
studies, the zone-folding method was firstly applied to study the
lattice dynamics of the SWCNT, which is migrated from that for the
planar graphite sheet. It successfully accounts the general
features of the phonon spectrum in SWCNT. However, unlike the
electronic problems, on which this method is workable except for
very narrow SWCNT, the results for the lattice vibration by the
zone-folding were not very satisfactory,\cite{Dresselhaus2}
particularly, it can not provide correct numbers of Raman and
infrared (IR) modes of SWCNT.\cite{Alon}

In recent a few years, the theoretical investigation on lattice
dynamics of SWCNT has achieved many successful and interesting
results. Using the rotational and helical symmetries, Popov {\it
et~al.}\cite{Popov1,Popov2} decomposed for the first time the
lattice dynamics equation of the SWCNT into a six-dimensional
eigenvalue problem. Analytical expression for the velocities of
the acoustic modes are derived using Born's perturbation
technique.\cite{Huang} The translational and rotational sum rules
expressed in terms of the force constant matrix\cite{Huang} of the
helical tube are also addressed in Ref.~\onlinecite{Popov2}.
Damnjanovic {\it et~al.} applied the line group analysis to the
lattice dynamics of SWCNT,\cite{Damnjanovic3,Damnjanovic4} and
also obtained the generalized Bloch form by modified group
projectors technique.\cite{Damnjanovic1, Damnjanovic8,
Damnjanovic2} They also calculated the phonon spectra in SWCNT
with a modified force constant model\cite{Damnjanovic5,
Damnjanovic6, Damnjanovic7} in which certain kinematic constraints
were imposed to achieve the twisting mode. The authors tried also
to fit their calculated frequencies of the Raman and IR modes as
functions of tube radius and chiral angel. In a series of papers
Mahan and his collaborators\cite{Mahan1, Mahan3} worked on the
lattice dynamics of the SWCNT with various vibrational energy
terms such as spring energy, in-plane or out-of-plane bond bending
potential energies successively, and calculated the phonon
spectrum for the achiral as well as chiral\cite{Ouyang} tubes.
They realized that, to keep the twisting mode to be a zero mode in
the long wave length limit, all the pieces of vibration energies
under their consideration violating the global rotational
invariance around axial axis should be ignored. The authors also
interpreted the existence condition of the flexure mode, which is
an inherent character for rods and plates. This mode also shows up
in Popov {\it et.~al.}'s calculation but is absent in Damnjanovic
{\it et~al.}'s results.

However, as pointed out in Ref.~\onlinecite{Mahan3}, the commonly
used expression for bond bending potential, even in addition with
the kinematic constraints as in Refs~\onlinecite{Damnjanovic5} and
\onlinecite{Damnjanovic6}, does not satisfy the rotational
invariance condition, where the missing of the flexure mode is a
kind of evidence. At the same time, the discussion on the
rotational symmetry requirements to the potential terms in
connection with the flexure mode is restricted only to those
around the tube axis. The force model\cite{Popov2} taken from that
for the graphite\cite{Aizawa} obeys the translational and
rotational sum rules of the planar sheet but does not meet
precisely the symmetry requirements of cylindrical warped
hexagonal lattice sheet.

Actually, the rotational invariance symmetry, i.e., the potential
energy must keep unchanged when a system is rigidly rotated,
should play a nontrivial role for rod-like cylindrical lattice
sheet. On the other hand, the intrinsic symmetry of the tube
lattice structure would also lead to specific restriction on
dynamic matrix. For a complete account of the symmetries of the
lattice dynamics for the SWCNT, both aspects should be considered
in a consistent way. This constitutes the basic scope of this
paper.

As a continuation of the previously mentioned series of works, in
this paper, we show the vibration potential, i.e. force model, up
in a detailed expression with the rigid translational and
rotational invariance being precisely guaranteed, in which the
curvature effect on bond lengths and bond angles is carefully
considered. As its direct consequences, the flexure mode as well
as the acoustic torsion mode can be correctly treated. Moreover we
calculate the phonon spectrum with the proposed force model and
carry out a full chiral symmetry analysis. Instead of previous
case-to-case empirical forms,\cite{Damnjanovic6} the frequencies
of all Raman and IR active modes, the polarization characteristics
of the modes, as well as the velocities of all four acoustic modes
can be systematically fitted in terms of tube radius and chiral
angle with the symmetry considerations.

As expected, our calculated results show that the curvature effect
on bond lengths and bond angles would be no more negligible only
for nanotubes  of smaller radius. Meanwhile the nontrivial
contribution for the $\theta$ dependence of varies physical
quantities is also effective only  for tubes with enough small
radius.  For thin tubes with the diameter range such as $4 \sim
10$~{\AA}, the warped graphene sheet model, on which our
discussions are based,  should be fairly applicable to the SWCNT.
For tubes with diameter smaller than 4~{\AA}  approximately, the
warped graphene sheet model might have to be improved by further
optimizing its equilibrium lattice configuration,\cite{Daniel,
Kurti, White2, Jackie} and the part of chiral symmetry analysis in
this paper were not be valid any more. Such estimation is entirely
consistent with the existing literatures.\cite{Damnjanovic6}

The present paper is organized as following. In Sec.~II, after a
brief description on notations, we present the detailed expression
of the vibrational potential. The rigid rotational invariance and
the comparisons with previous works are discussed. Subsection
III~A is devoted to a generic chiral symmetry analysis for various
physical quantities, while the main results  and relevant
discussions on phonon spectrum calculations are presented in
Subsec.~III~B. The paper ends with a brief summary in Sec.~IV.

\section{MODEL}
\subsection{Notations}
\begin{figure}
    \begin{center}
        \scalebox{0.6}[0.6]{\includegraphics[width=7cm]{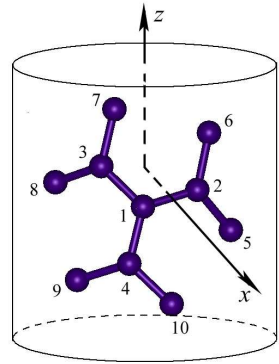}}
    \end{center}
    \caption{The sketch of carbon atoms on a cylindric surface.}
    \label{fig:nanotube}
\end{figure}

The $z$ axis in this paper is set along the tube axis and the $x$
axis comes across the middle point of a \mbox{C-C} bond in
Fig.~\ref{fig:nanotube}. It is understood in this subsection that
notations for cylindrically warped tubes are often introduced with
the aid of those defined on the planar sheet.

As is well known, SWCNT can be notated by a chiral vector
$\vec{R}=n_{1}\vec{a}_{1}+n_{2}\vec{a}_{2}$ or equivalently, the
radius $r$ and chiral angle $\theta$. Where $\vec{a}_{1}$ and
$\vec{a}_{2}$ being the primitive lattice vectors in graphite
lattices,\cite{Dresselhaus2}
$|\vec{a}_{1}|=|\vec{a}_{2}|=2.46$~{\AA}. They have the relations
as
\begin{eqnarray}
    r=\frac{|\vec{a}_{1}|}{2\pi}\sqrt{n_{1}^{2}+n_{1}n_{2}+n_{2}^{2}},\hspace{0.5cm}
    \theta=\mbox{atan}\frac{\sqrt{3}n_{2}}{2n_{1}+n_{2}}\;.
    \label{chiralangle}
\end{eqnarray}

We use the helical and rotational symmetry descriptions for the
SWCNT throughout this paper. Introduce a real lattice vector
$\vec{H}=p_{1}\vec{a}_{1}+p_{2}\vec{a}_{2}$ describing the screw
operation\cite{White} with $n_{1}p_{2}-n_{2}p_{1}=N$ ($N$ is the
greatest common divisor of $n_{1}$ and $n_{2}$). The rotational
angle of the screw operation is
$\alpha=(\vec{H}\cdot\vec{R})/|\vec{R}|^{2}$  (in the unit of
$2\pi$). A unit cell in the $(n_{1}, n_{2})$ tubule can be notated
by two equivalent ways, either based on $(\vec{H},
\frac{\vec{R}}{N})$, or on $(\vec{a}_1, \vec{a}_2)$ as
\begin{eqnarray}
      \vec{r}_{m,l}=m\vec{H}+l\frac{\vec{R}}{N}, \hspace{0.25cm}
      \mbox{or} \hspace{0.25cm}
      \vec{r}_{q_{1},q_{2}}=q_{1}\vec{a}_{1}+q_{2}\vec{a}_{2}
     \label{whitenotation}
\end{eqnarray}
with the relations
\begin{eqnarray}
     m=(n_{1}q_{2}-n_{2}q_{1})/N, \hspace{0.25cm}
     l=q_{1}p_{2}-q_{2}p_{1}\;.
     \label{whitelabel}
\end{eqnarray}
In the following, we use $(\vec{H}, \frac{\vec{R}}{N})$ as the
basic vectors, and $\vec{b}_{H}$ and $\vec{b}_{R}$ are their
corresponding reciprocal unit vectors. Any wave vector in the
reciprocal space can be written as
\begin{eqnarray}
\vec{K}=\frac{\kappa}{2\pi}\vec{b}_{H}+\frac{n}{N}\vec{b}_{R}\;.
\end{eqnarray}
In the first Brillouin zone, $\kappa\in(-\frac{1}{2},
\frac{1}{2}]$ and $n$ is an integer in $(-\frac{N}{2},
\frac{N}{2}]$.

\subsection{Potentials}
It had been proposed\cite{Aizawa} that there are five distinctive
terms for the vibrational potential of the graphene sheet
satisfying the rigid translational and rotational invariance of
the planar sheet. We herein present a detailed expression of the
vibrational potential for the SWCNT with the curvature effect
being carefully in-cooperated, which is essentially the
generalization of that for the graphene sheet. It satisfies
precisely the requirements of the rigid translational and
rotational invariance and realizes the corresponding general
symmetry sum rules in Ref.~\onlinecite{Popov2}, i.e., the
potential energy must keep vanish term by term when the tube is
rigidly translated or rotated around any axis. Introduce
$\vec{r}_{i}$ as the equilibrium position of atom \textit{i} and
$\vec{u}_{i}$ as its displacement vector.
$\vec{r}_{ij}=\vec{r}_j-\vec{r}_i$ is the vector from atom
\textit{i} to \textit{j} in the nanotube while the modulus
$r_{ij}$ represents the length of \mbox{C-C} bond between atoms
\textit{i} and \textit{j}. $\vec{r}_{i}$ is determined following
the geometry of a warped graphene sheet so that the three
tridental bonds nearest-neighbored with the atom \textit{i} as
well as the angles between any of the two bonds are not equal to
each other, especially for thin tubes.

The five potential terms for the SWCNT are expressed in the
following.

(1) $V_{l}$ is the potential of the spring force between the
nearest-neighbor atom pair,
\begin{eqnarray}
    V_{l}=\frac{k_{l}}{2}\sum_{i=2}^{4}
    [(\vec{u}_{i}-\vec{u}_{1})\cdot\vec{e}_{1i}^{l}]^{2}\;,
    \label{Potential1}
\end{eqnarray}
where $k_{l}$ is the first-order force constant and $
\vec{e}_{1i}^{l}=\frac{\vec{r}_{1i}}{|\vec{r}_{1i}|}$. We'd like
to point out that the components of the displacement vectors
perpendicular to $\vec{e}_{1i}^{l}$ which violates  the rigid
rotational invariance are forbidden.

(2) $V_{sl}$ is also the potential of the spring force but between
the next nearest-neighbored atoms illustrated as  $(1, \, 5 \ldots
10)$ in Fig.~\ref{fig:nanotube},
\begin{eqnarray}
    V_{sl}=\frac{k_{sl}}{2}\sum_{i=5}^{10}
    [(\vec{u}_{i}-\vec{u}_{1})\cdot\vec{e}_{1i}^{l}]^{2}
    \label{Potential2}
\end{eqnarray}
with $k_{sl}$  the second-order force constant.

(3) The potential energy for the in-surface bond bending $V_{BB}$
is actually a term associated with bond angle variations. Three
atoms have to be considered together, and
\begin{eqnarray}
    V_{BB}
    &=&\frac{k_{BB}}{4}\sum_{j_i}
    \sum_{{\scriptstyle j'_i \atop \scriptstyle (j'_i\neq j_i)}}\left[\frac{\vec{u}_{j_i}-\vec{u}_{i}}{r_{ij_i}}
    \cdot(\vec{e}_{ij'_i}^{l}-\cos\theta_{j_iij'_i}\vec{e}_{ij_i}^{l})\right.\nonumber\\
    & &\left.+\frac{\vec{u}_{j'_i}-\vec{u}_{i}}{r_{ij'}}\cdot(\vec{e}_{ij_i}^{l}-\cos\theta_{j'_iij_i}\vec{e}_{ij'_i}^{l})\right]^{2}\nonumber\\
    &=&\frac{k_{BB}}{4}\sum_{j_i}
    \sum_{{\scriptstyle j'_i \atop \scriptstyle (j'_i\neq j_i)}}
    (\cos\theta'_{j_iij'_i}-\cos\theta_{j_iij'_i})^{2}\;.
    \label{Potential3}
\end{eqnarray}
If $i$ is 1 or 2, $j_i, j'_i$ take the sites $2,3,4$ or $1,5,6$
respectively.  In Eq.~(\ref{Potential3}), $\theta_{j_iij'_i}$
stands for the equilibrium angle between the bonds
$\vec{r}_{ij_i}$ and $\vec{r}_{ij'_i}$, while $\theta'_{j_iij'_i}$
for the corresponding angle in vibration. The expression in terms
of $\vec{u}_{j_i}-\vec{u}_{i}$ and $\vec{u}_{j'_i}-\vec{u}_{i}$ is
an honest representation to that of $\cos{\theta'}$, where the
rigid rotational invariance referred to an arbitrary axis can be
kept only when the differences among bond lengths and bond angles
be carefully accounted.

(4) The potential of the out-of-surface bond bending $V_{rc}$ is
the energy between atom $i$ and the three nearest-neighbor atoms
$j_i$ with four atoms simultaneously contained,
\begin{equation}
    V_{rc}=\frac{k_{rc}}{2}[(3\vec{u}_{i}-\sum_{j_{i}}\vec{u}_{j_{i}})
    \cdot\vec{e}_{i}^{rc}]^{2},
    \label{Potential4}
\end{equation}
\begin{equation}
\vec{e}_{i}^{rc}=-\frac{\sum_{j_{i}}\vec{r}_{j_{i}}}{|\sum_{j_{i}}\vec{r}_{j_{i}}|}\;,
\label{newvector}
\end{equation}
Where $i$ takes 1 or 2 with $j_i$ running over the three nearest
neighbors of atom $i$. This potential has the physical intuition
as that responsible for the radial optical mode in tubes and is
trying to keep the four atoms on the cylinder surface. Herein we
introduce an unit vector $\vec{e}_{i}^{rc}$ defined  in
Eq.~(\ref{newvector}), which is rather close but not equal to the
radial unit vector of site 1 shown in Fig.~2. When the radius is
large enough, the $\vec{e}_{i}^{rc}$ would be close enough to the
$\vec{e}_{1}^{r}$. However, we stress that the potential term
$V_{rc}$ with $\vec{e}_{i}^{rc}$ keeps the rigid rotational
invariance while that with $\vec{e}_{i}^{rc}$ substituted by
$\vec{e}_{i}^{r}$ would break the rotational symmetry.
\begin{figure}[htpb]
    \begin{center}
        \scalebox{0.7}[0.7]{\includegraphics{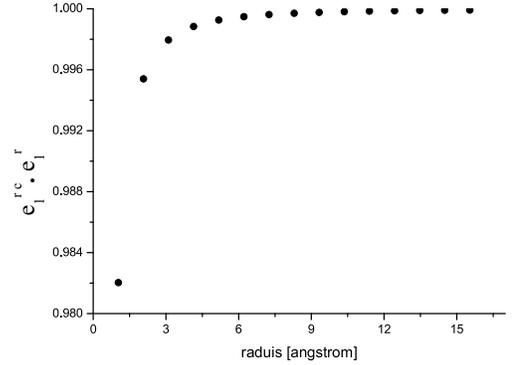}}
    \end{center}
    \caption{The projection of $\vec{e}_{1}^{rc}$ on $\vec{e}_{1}^{r}$ for
             tubes $(2n, n)$ with $n\in[1, 15]$.
             It shows that $\vec{e}_{1}^{rc}$ only deviates about 2\% from $\vec{e}_{1}^{r}$
            even in the small radius \mbox{(2, 1)} tube.}
    \label{fig:E1rc}
\end{figure}

(5) The twist potential energy for bond $\vec{r}_{1k}$ is
\begin{eqnarray}
    V_{tw}=\frac{k_{tw}}{2}\sum_{<i,j>}
    [(\vec{u}_{i}-\vec{u}_{j}-(\vec{u}_{i'}-\vec{u}_{j'}))\cdot\vec{e}_{1k}^{r}]^{2}\;,
    \label{Potential5}
\end{eqnarray}
where $\vec{e}_{1k}^{r}$ is the unit vector along the radial
direction of the middle point of $\vec{r}_{1k}$,  $<i,j>$
represents a pair of atoms nearest-neighbored with atom 1 while
$k$ the third of its nearest neighbors. Pair $<i',j'>$ is the
image of $<i,j>$ referring to a $C_{2}$ rotation around the axis
in $\vec{e}_{1k}^{r}$. It has the intuition that responsible for
modes with twisted vibrations.

\subsection{Some details about the model}
Obviously, all the five potential energy terms respect the
translational invariance. When $\vec{u}_{i}=\vec{u}_{j}$,
$\vec{u}_{i}-\vec{u}_{j}=0$ leads to
$$
    V_{l}=V_{sl}=V_{BB}=V_{rc}=V_{tw}=0 \;.
    \label{translational}
$$

For the rotational invariance, we have to consider the five
potentials term by term separately.\cite{Huang, Madelung} When the
tube rotates rigidly around an arbitrary axis for a small angle
$\delta\vec{\omega}$ with its direction along the axis
$\frac{\delta\vec{\omega}}{|\delta\vec{\omega}|}$ and its
magnitude $|\delta\vec{\omega}|$ being the totation angle, each
lattice site acquires a displacement
$\vec{u}_{i}=\delta\vec{\omega}\times\vec{r}_{i}$,
\begin{eqnarray}
 \vec{u}_{i}-\vec{u}_{j}=\delta\vec{\omega}\times(\vec{r}_{i}-\vec{r}_{j})
 =\delta\vec{\omega}\times\vec{r}_{ji}\;.
\label{rotational}
\end{eqnarray}
Substituting Eq.~(\ref{rotational}) into the first two potential
terms (\ref{Potential1}) and (\ref{Potential2}), it is
straightforward to have
$(\vec{u}_{j}-\vec{u}_{i})\cdot\vec{e}_{ij}^{l}
=r_{ij}(\delta\vec{\omega}\times\vec{e}_{ij}^{l})\cdot\vec{e}_{ij}^{l}
=0.$ Then
$$
 V_{l}=V_{sl}=0 \; .
$$
Substituting Eq.~(\ref{rotational}) into the third potential term
(\ref{Potential3}), a typical representative term in summation
becomes
$$
 V_{BB}\sim\frac{k_{BB}}{4}[\delta\vec{\omega}\cdot(\vec{e}_{12}^{l}\times\vec{e}_{13}^{l}
 +\vec{e}_{13}^{l}\times\vec{e}_{12}^{l})]^{2}=0\;.
$$
In which a fact has been used that $r_{ij}$ in the denominate is
cancelled by that in the numerator when Eq.~(\ref{rotational}) is
applied. Moreover, for each typical representative term in
potentials (\ref{Potential4}) and (\ref{Potential5}), we have
similarly,
\begin{eqnarray*}
    \begin{array}{l}
    V_{rc}\sim\frac{k_{rc}}{2}[\delta\vec{\omega}\times(\vec{r}_{12}
    +\vec{r}_{13}+\vec{r}_{14})\cdot\vec{e}_{1}^{rc}]^{2}=0\; ,\\
    V_{tw}\sim\frac{k_{tw}}{2}[\delta\vec{\omega}\times(\vec{r}_{43}
    -\vec{r}_{56})\cdot\vec{e}_{12}^{r}]^{2}=0\;.
    \label{rotational2}
    \end{array}
\end{eqnarray*}

\begin{figure}[htpb]
    \begin{center}
        \scalebox{0.8}[0.8]{\includegraphics[width=8.5cm]{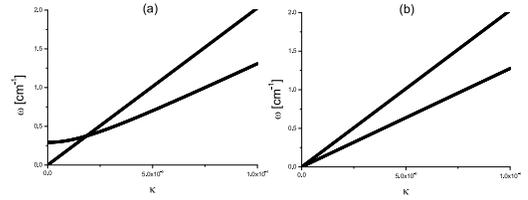}}
    \end{center}
    \caption{The effect of bond lengths on TW mode in tube (5, 2).
     (a) All bonds are assumed to
    be the same. The frequency of TW mode is nonzero.
     (b) The differences between bonds
    are considered. The frequency of TW mode is presicely zero.}
    \label{fig:twistingmode}
\end{figure}

As an example for further clarification, one might calculate the
phonon spectrum with all the bond lengths and bond angles assumed
to be equal to that of the graphite. It results that the twisting
mode (TW) at $(\kappa, n)=(0, 0)$ is no longer a zero mode  and
there is a finite gap. For the SWCNT (5,2), the error introduced
by the equal-bond-length creates a gap of the order
$0.5$~cm$^{-1}$ as shown in Fig.~\ref{fig:twistingmode},  where
the radius $r=2.45$~{\AA}, the three \mbox{C-C} bonds are about
$-1.3$\%, $-0.3$\% and $0.0$\% shorter than that of graphite
$1.42$~{\AA} respectively. Although it is a minute number and
entirely negligible in practice, it is of qualitative
significance.

As a distinguished feature of the nanotubes that the two
degenerate transversal acoustic (TA) modes shown up at
$(\kappa,n)=\pm(\alpha,1)$ are flexure modes. Instead of the
conventional linear behaviors, the low frequency limits of their
dispersions are parabolic as
$\omega^{2}=\beta^{2}(\kappa\mp\alpha)^{4}$. We address that the
rigid rotational invariance around $z$ axis itself is not a
sufficient condition for the existence of the flexure modes cf.
Ref.~\onlinecite{Mahan3}. As a counter example, we introduce a
term $ V_{\tau i}=\frac{k_{\tau
i}}{2}[(\vec{u}_{2}-\vec{u}_{1})\cdot\vec{e}_{12}^{\tau i}]^{2}$
with $\vec{e}_{12}^{\tau
i}=\vec{e}_{12}^{r}\times\vec{e}_{12}^{l}$, which is the elastic
energy due to the tangential spring force between atoms 1 and 2.
It is easy to verify that $V_{\tau i}=0$ when the tube rotates
around the $z$ axis. But there is no flexure mode when $V_{\tau
i}$ is taken into account.\cite{Damnjanovic6} This is because that
$V_{\tau i}$ would be no more zero when the tube  rotates around
any axis perpendicular to the $z$ axis.

\section{RESULTS AND DISCUSSIONS}
\subsection{Further symmetry considerations}
We may establish a mapping from the space of chiral vectors on the
planar sheet to that of the nanotube structure in a fixed way of
wrapping\cite{Ye2}
\begin{eqnarray*}
    f:\{\mbox{chiral vector set} \}\mapsto\{\mbox{nanotube set}
    \}\;.
\end{eqnarray*}
However this is not a one-to-one mapping. As mentioned in
Ref.~\onlinecite{Ye2}, a given SWCNT can be equivalently composed
by three chiral vectors $\vec{R}_{0}$, $\vec{R}_2$ and $\vec{R}_4$
with chiral angles as $\theta$, $\theta+2\pi/3$ and
$\theta+4\pi/3$ respectively within the corresponding graphite
sheet. Due to the $\hat{C}_{6v}$ symmetry of the hexagonal
graphene lattice sheet, there is another set of vectors
$\vec{R}_1$, $\vec{R}_3$ and $\vec{R}_5$ with chiral angles as
$\theta+\pi/3$, $\theta+\pi$ and $\theta+5\pi/3$ respectively on
the graphite sheet. These three chiral vectors correspond again to
the same SWCNT which is actually the nanotube by rotating the tube
formed by $\vec{R}_0$ upside down, i.e. a $\hat{C}_{2x}$
operation. The net result of the operation is an exchange of A and
B carbon atoms in unit cells with the sign of its chiral index
$\nu=\mbox{mod}\{n_1-n_2, 3\}$ also changed. There is one another
operation $\hat{\sigma}_{xz}$ connecting a pair of SWCNT which is
the mirror reflection onto each other with respect to $xz$ plane.
Correspondingly  $\vec{R}_0$ in the sheet is changed into its
countpart $\vec{R}_{0}^{'}$ with $\theta \rightarrow -\theta$ but
$\nu$ kept unchanged.

As one of the direct consequences of the above observations, any
physical quantity $Q^{\nu}(r,\theta)$ of the SWCNT is a periodical
function of the chiral angle $\theta$ with period
$\frac{2\pi}{3}$, i.e., $Q^{\nu}(r,\theta)$ can be expanded
as\cite{Ye2}
\begin{eqnarray}
    Q^{\nu}(r,\theta)=\sum_{n=0}^{\infty}a_{n}^{\nu}\cos(3n\theta)
    +b_{n}^{\nu}\sin(3n\theta)\;.
    \label{Fourierexpansion}
\end{eqnarray}

Furthermore, under the operation
$\theta\rightarrow\theta+{\pi}/{3}$, any scalars $S$ would keep
unchanged, For normal vectors, the radial components have the same
behavior, but the azimuthal and axial components change
signs\cite{Ye2}. That is
\begin{equation}\left\{\begin{array}{l}
    S^{\nu}(\theta+\frac{\pi}{3})=S^{-\nu}(\theta)\;,
    \\
    \vec{v}^{\nu}(\theta+\frac{\pi}{3})=\hat{C}_{2x}\vec{v}^{-\nu}(\theta)
\end{array} \right. \label{C2x}\end{equation}
with $\hat{C}_{2x}\vec{e}_r=\vec{e}_r$,
$\hat{C}_{2x}\vec{e}_{\phi}=-\vec{e}_{\phi}$,
$\hat{C}_{2x}\vec{e}_z=-\vec{e}_z$. Here $\vec{e}_r$,
$\vec{e}_\phi$ and $\vec{e}_z$ are unit vectors oriented towards
radial, azimuthal and axial directions upon the cylindrical
surface respectively.

In addition, operation $\hat{\sigma}_{xz}$ gives that any scalars
or radial and axial components of vectors are even functions of
$\theta$, while azimuthal components are odd functions of
$\theta$,\cite{Ye2}
\begin{equation}\left\{\begin{array}{l}
      S^{\nu}(-\theta)=S^{\nu}(\theta)\;,    \\
    \vec{v}^{\nu}(-\theta)=\hat{\sigma}_{xz}\vec{v}^{\nu}(\theta)
\end{array} \right. \label{sigmaxz}\end{equation}
 with $\hat{\sigma}_{xz}\vec{e}_r=\vec{e}_r$,
$\hat{\sigma}_{xz}\vec{e}_{\phi}=-\vec{e}_{\phi}$,
$\hat{\sigma}_{xz}\vec{e}_z=\vec{e}_z$. The detailed expressions
reduced from Eq.~(\ref{Fourierexpansion}) due to the symmetry
restrictions Eqs~(\ref{C2x}) and (\ref{sigmaxz}) are shown in the
appendix.

\subsection{Frequencies, acoustic velocities, and eigenvectors}
We calculated the phonon spectrum in the $(\kappa,n)$
representation with the vibrational potential
Eqs~(\ref{Potential1})-(\ref{Potential5}). By tuning the
calculated results to the experimental data\cite{Dresselhaus4} for
the Raman modes of (10, 10) tube (see
Table~\ref{Tab:armchair1010}), we fit the corresponding force
constants as $k_{l}=364.0$~Nm$^{-1}$, $k_{sl}=62.0$~Nm$^{-1}$,
$k_{BB}=1.07\times10^{-11}$~erg, $k_{rc}=14.8$~Nm$^{-1}$, and
$k_{tw}=6.24$~Nm$^{-1}$. We successfully identified 4 zero modes,
14 Raman active modes and 6 IR active modes for the chiral
SWCNT.\cite{Alon} Two of the six modes at $(\kappa, n)=(0, 0)$ are
zero modes. They are longitudinal acoustic (LA) and TW modes both
belonging to $_{0}A_{0}^{-}$ representation (Reps). Another mode
belonging also to $_{0}A_{0}^{-}$ Reps is the IR active
$\vec{e}_{r}$ optical (OP) mode with A and B atoms oscillating out
of surface in radial direction. The other three nonzero modes at
$(\kappa,n)=(0,0)$ belonging to $_{0}A_{0}^{+}$ Reps are Raman
active, they are $\vec{e}_r$ acoustic (AC), $\vec{e}_{\theta}$ OP
and $\vec{e}_{z}$ OP. There are one flexure and five nonzero modes
at $(\kappa, n)=(\alpha, 1)$. The latter are both Raman and IR
active assigned to $_{\alpha}E_{1}$ Reps. At $(\kappa,
n)=(2\alpha, 2)$, six nonzero modes are all Raman active belonging
to the same $_{2\alpha}E_{2}$ Reps.

\begin{table*}[t]
     \caption{Comparison between the calculated results
     (following our model) and the experimental values
     for several mode frequencies (in the unit of cm$^{-1}$)
     of SWCNT (10,10).}
     \label{Tab:armchair1010}
\begin{ruledtabular}
\begin{tabular}{cllllllll}
Reps                  & $_{0}A_{0}^{+}$       & $_{0}A_{0}^{+}$       &
$_{\alpha}E_{1}^{-}$  & $_{\alpha}E_{1}^{-}$  & $_{2\alpha}E_{2}^{+}$ &
$_{2\alpha}E_{2}^{+}$ & $_{2\alpha}E_{2}^{+}$ & $_{2\alpha}E_{2}^{+}$ \\
\hline
theory & 167 & 1588 & 105 & 1588 & 21 & 367 & 873 & 1584\\
experiment\cite{Dresselhaus4} & 186 & 1593 & 118 & 1567 & / & 377 & / & 1606\\
\end{tabular}
\end{ruledtabular}
\end{table*}

\begin{figure}[htpb]
    \begin{center}
        \scalebox{1.4}[1.4]{\includegraphics[width=4cm]{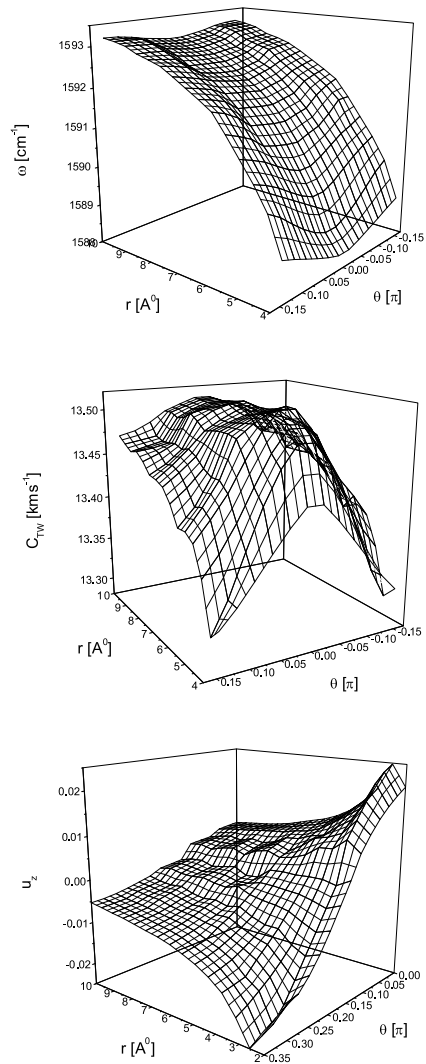}}
    \end{center}
    \caption{As functions of $r$ and $\theta$, we show in
    (a) the frequency of $\vec{e}_{z}$ OP mode at $(\kappa, n)=(0,0)$;
    (b) $C_{\mbox{\tiny TW}}$ for small radius tubes;
    (c) the $z$ component of the eigenvector of $\vec{e}_{r}$ AC
    mode. }
    \label{fig:phyrtheta}
\end{figure}

We also calculated the sound velocities of LA and TW modes, the
flexure parabolic coefficient for the TA mode, and polarizations
of nonzero modes at $(\kappa,n)=(0,0)$ for varieties of tubes as
the functions of $r$ and $\theta$. The modes at
$(\kappa,n)=(\alpha,1)$ as well as $(2\alpha,2)$ are doubly
degenerated with their one-to-one degenerate counterparts at
$(-\alpha,-1)$ and $(-2\alpha,-2)$ respectively. In
Fig.~\ref{fig:phyrtheta}, we provide the three-dimensional plots
for some of these quantities as functions of $r$ and $\theta$.

\renewcommand{\arraystretch}{1.5}
\begin{table*}[t]
\caption{Frequencies (in the unit of cm$^{-1}$) of 15 Raman and IR
active modes as functions of $r$ (in \AA) and $\theta$.}
\label{Tab:omega}
\begin{ruledtabular}
\begin{tabular}{crll}
Reps & Mode/Order &  $\omega(\theta)$ &   $f_{i}(r)$   \\
\hline
$_{0}A_{0}^{+}$ R& $\vec{e}_{r}$ AC, 1 & $f_{0}(r)$ & $f_{0}(r)=\frac{1133.86}{r}-\frac{139.65}{r^{3}}$\\
  &$\vec{e}_{\theta}$ OP, 2 & $f_{0}(r)+f_{1}(r)\cos6\theta$ & $f_{0}(r)=1594.00-\frac{266.98}{r^{2}}$, $f_{1}(r)=\frac{8.65}{r^{2}}$\\
  &$\vec{e}_{z}$ OP, 3 & $f_{0}(r)+f_{1}(r)\cos6\theta+f_{2}(r)\cos12\theta$ & $f_{0}(r)=1594.00-\frac{91.81}{r^{2}}$,
  $f_{1}(r)=-\frac{15.68}{r^{2}}$, $f_{2}(r)=-\frac{0.68}{r^{2}}$  \\
$_{0}A_{0}^{-}$ Ir&$\vec{e}_{r}$ OP, 4 & $f_{0}(r)+f_{1}(r)\cos6\theta$ & $f_{0}(r)=864.81+\frac{990.22}{r^{2}}-\frac{1117.30}{r^{4}}$, $f_{1}(r)=\frac{9.16}{r^{2}}$\\
\hline
$_{\alpha}E_{1}$ R  Ir & 1 & $f_{0}(r)+f_{1}(r)\cos6\theta$ & $f_{0}(r)=\frac{710.16}{r}+\frac{45.07}{r^{3}}$, $f_{1}(r)=\frac{1.99}{r^{3}}-\frac{31.55}{r^{4}}$\\
  & 2 & $f_{0}(r)+f_{1}(r)\cos6\theta$ & $f_{0}(r)=\frac{1603.51}{r}-\frac{746.51}{r^{3}}$, $f_{1}(r)=-\frac{115.54}{r^{3}}$\\
  & 3 & $f_{0}(r)+f_{1}(r)\cos6\theta$ & $f_{0}(r)=864.84+\frac{860.00}{r^{2}}-\frac{1758.70}{r^{4}}$, $f_{1}(r)=\frac{11.63}{r^{2}}-\frac{206.52}{r^{4}}$\\
  & $\nu=\pm 1$, 4 & $f_{0}(r) \pm f_{1}(r)\cos9\theta$ & $f_{0}(r)=1594.13-\frac{316.67}{r^{2}}$, $f_{1}(r)=\frac{31.92}{r^{3}}$\\
  & $\nu=0$, 4 & $f_{0}(r)+f_{1}(r)\cos6\theta+f_{2}(r)\cos12\theta$ & $f_{0}(r)=1594.14-\frac{318.48}{r^{2}}$, $f_{1}(r)=\frac{7.83}{r^{2}}-\frac{19.03}{r^{4}}$\\
  &&&\hspace{1cm}$f_{2}(r)=\frac{2.70}{r^{2}}+\frac{0.60}{r^{4}}$\\
  & 5 & $f_{0}(r)+f_{1}(r)\cos6\theta$ & $f_{0}(r)=1593.97-\frac{277.49}{r^{2}}$, $f_{1}(r)=-\frac{12.45}{r^{2}}$\\
\hline
$_{2\alpha}E_{2}$ R & 1 & $f_{0}(r)+f_{1}(r)\cos6\theta+f_{2}(r)\cos12\theta$ & $f_{0}(r)=\frac{959.33}{r^{2}}-\frac{736.60}{r^{4}}+\frac{779.59}{r^{5}}$\\
&&&\hspace{1cm} $f_{1}(r)=\frac{6.19}{r^{3}}+\frac{73.37}{r^{4}}$, $f_{2}(r)=-\frac{0.06}{r^{3}}+\frac{9.34}{r^{4}}$\\
  & 2 & $f_{0}(r)+f_{1}(r)\cos6\theta$ & $f_{0}(r)=\frac{1420.21}{r}+\frac{54.52}{r^{3}}-\frac{1246.29}{r^{5}}$, $f_{1}(r)=\frac{204.34}{r^{3}}$ \\
  & 3 & $f_{0}(r)+f_{1}(r)\cos6\theta$ & $f_{0}(r)=\frac{2535.48}{r}-\frac{2426.65}{r^{3}}$, $f_{1}(r)=-\frac{412.23}{r^{3}}$ \\
  & 4 & $f_{0}(r)+f_{1}(r)\cos6\theta$ & $f_{0}(r)=864.80+\frac{486.71}{r^{2}}-\frac{4711.81}{r^{4}}+\frac{12425.61}{r^{6}}$\\
&&&\hspace{1cm} $f_{1}(r)=\frac{9.89}{r^{2}}-\frac{524.74}{r^{4}}$\\
  & 5 & $f_{0}(r)+f_{1}(r)\cos6\theta$ & $f_{0}(r)=1594.00-\frac{869.19}{r^{2}}+\frac{978.77}{r^{4}}$\\
&&&\hspace{1cm} $f_{1}(r)=-\frac{16.15}{r^{2}}+\frac{363.41}{r^{4}}$\\
  & 6 & $f_{0}(r)+f_{1}(r)\cos6\theta$ & $f_{0}(r)=1594.01-\frac{392.92}{r^{2}}-\frac{2160.15}{r^{4}}+\frac{5416.26}{r^{6}}$\\
&&&\hspace{1cm} $f_{1}(r)=\frac{7.88}{r^{2}}-\frac{297.88}{r^{4}}$\\
\end{tabular}
\end{ruledtabular}
\end{table*}

Our calculated results are further fitted following the
parameterization shown in  Eq.~(\ref{Fourierexpansion}). The
corresponding expressions for the frequencies of above 15 Raman
and IR active modes, the velocities and flexure parabolic, and the
polarization vectors of nonzero modes at $(\kappa,n)=(0,0)$  are
listed in Tables~\ref{Tab:omega}, \ref{Tab:velocity}, and
\ref{Tab:eigenvector} respectively. In this paper particular
emphasis is drown to tubes with smaller radius. So we take  the
parameter regime for frequencies and velocities in $r\in[4.0,
10.0]$~{\AA} and $\theta\in[-\frac{\pi}{6}, \frac{\pi}{6}]$, while
that for polarization vectors in $r\in[2.0, 10.0]$~{\AA} and
$\theta\in[-\frac{\pi}{3}, \frac{\pi}{3}]$. We keep the relative
errors in fitting less than $5 \times 10^{-4}$. The above
parameter regimes are such chosen only for tubes of smaller
radius, in which the contributions of $\theta$ dependence are
notable comparing to those of $r$ dependence. The velocity of the
TW mode (in Table~\ref{Tab:velocity}) as well as the polarization
vectors of modes with $_0A_0^{+}$ Reps (in
Table~\ref{Tab:eigenvector}) are typical examples shown up an
evident $\theta$ dependence. In fact, the corresponding
calculations have also been carried out in the parameter regime
$r\in[3.13, 58.7]$~{\AA}. The results can be fitted very well by
the same functions  with minor change of coefficients, i.e., our
fitting formulae could be applied to a much wider parameter
regime.

\begin{table*}[t]
\caption{Sound velocities (in kms$^{-1}$) of the TW and LA modes,
and $\beta $ (in $10^{-6}$m$^{2}$s$^{-1}$) of the flexure mode as
functions of $r$ (in {\AA}) and $\theta $.} \label{Tab:velocity}
\begin{ruledtabular}
\begin{tabular}{cll}
Velocity & Velocity($\theta$) & $f_{i}(r)$ \\
\hline
$C_{\mbox{\scriptsize TW}}$ & $f_{0}(r)+f_{1}(r)\cos6\theta$ & $f_{0}=13.5-\frac{1.63}{r^{2}}$, $f_{1}=\frac{2.38}{r^{2}}$\\
$C_{\mbox{\scriptsize LA}}$ & $f_{0}(r)+f_{1}(r)\cos6\theta$ & $f_{0}=21.0706+\frac{0.0055}{r}-\frac{0.6860}{r^{2}}$, $f_{1}=\frac{0.00091}{r}-\frac{0.01679}{r^{2}}$\\
$\beta$ & $f_{0}(r)+f_{1}(r)\cos6\theta$ & $f_{0}=1.3767r-0.00142r^{2}-5.8\times10^{-5}r^{3}$, $f_{1}=-\frac{0.143}{r}+\frac{0.04994}{r^{3}}$\\
\end{tabular}
\end{ruledtabular}
\end{table*}

\renewcommand{\arraystretch}{1.5}
\begin{table*}[t]
\caption{Polarization vectors
$\vec{u}\equiv(\vec{u}(A),\vec{u}(B))$ as functions of $r$ (in
\AA) and $\theta$. Where $\vec{u}(A)$ and $\vec{u}(B)$ indicate
the displacement vectors of  atoms A and B in the $(0, 0)$ unit
cell respectively. For the three modes in this table,
$u_{r}(B)=u_{r}(A)$, $u_{\phi}(B)=-u_{\phi}(A)$,
$u_{z}(B)=-u_{z}(A)$. } \label{Tab:eigenvector}
\begin{ruledtabular}
\begin{tabular}{ccll}
$(\kappa, n)=(0, 0)$ & & Vector($\theta$) & $f_{i}(r)$  \\
\hline
$R_{1}$ ($\vec{e}_{r}$ AC) & $u_{r}(A)$ & $f_{0}(r)$ & $f_{0}(r)=0.7071-\frac{0.0028}{r^{2}}$ \\
& $u_{\phi}(A)$ & $f_{1}(r)\sin3\theta$ & $f_{1}(r)=\frac{0.0518}{r}+\frac{0.0468}{r^{2}}$ \\
& $u_{z}(A)$ & $f_{1}(r)\cos3\theta$ & $f_{1}(r)=\frac{0.0517}{r}+\frac{0.0749}{r^{2}}$ \\
\hline
$R_{2}$ ($\vec{e}_{\phi}$ OP) & $u_{r}(A)$  & $f_{1}(r)\sin3\theta$ & $f_{1}(r)=-\frac{0.0542}{r}-\frac{0.0455}{r^{2}}$ \\
& $u_{\phi}(A)$ & $f_{0}(r)+f_{1}(r)\cos12\theta$ & $f_{0}(r)=0.7056+\frac{0.0019}{r^{2}}$, $f_{1}(r)=0.0015-\frac{0.003}{r^{2}}$\\
& $u_{z}(A)$ & $f_{1}(r)\sin6\theta+f_{2}(r)\sin12\theta$ & $f_{1}(r)=0.0656-\frac{0.0801}{r^{2}}$, $f_{2}(r)=0.0048-\frac{0.0112}{r^{2}}$ \\
\hline
$R_{3}$ ($\vec{e}_{z}$ OP) & $u_{r}(A)$ & $f_{1}(r)\cos3\theta$ & $f_{1}(r)=-\frac{0.0447}{r}-\frac{0.0417}{r^{2}}$\\
& $u_{\phi}(A)$ & $f_{1}(r)\sin6\theta+f_{2}(r)\sin12\theta$ & $f_{1}(r)=-0.0656+\frac{0.0773}{r^{2}}$, $f_{2}(r)=-0.0048+\frac{0.0111}{r^{2}}$\\
& $u_{z}(A)$ & $f_{0}(r)+f_{1}(r)\cos12\theta$ & $f_{0}(r)=0.7056+\frac{0.0019}{r^{2}}$, $f_{1}(r)=0.0015-\frac{0.0033}{r^{2}}$\\
\end{tabular}
\end{ruledtabular}
\end{table*}

It is interesting to see that all the numerically fitting
expressions meet the symmetry requirements of Eqs~(\ref{C2x}) and
(\ref{sigmaxz}), where the velocities and frequencies are of
scalars while polarizations are of vectors. We notice that in
Table~\ref{Tab:omega}, one of the $_{\alpha}E_{1}$ modes manifests
different parameter dependence for different chiral index $\nu$,
i.e. it has different expressions for $\nu=\pm 1$ and $\nu=0$
respectively. However, it is still consistent with the general
expressions Eqs~(\ref{C2x}) and (\ref{sigmaxz}). We further notice
that, different from those physical quantities of normal vectors,
the polarization vector can be measured up to a global phase
factor as $\pm 1$. Therefore, the corresponding transformation
properties with respect to $\theta\rightarrow
\theta+\frac{\pi}{3}$ as well as $\theta\rightarrow -\theta$ need
to be generalized to
\begin{equation}\left\{\begin{array}{l}
v_P^{\nu(m)}(\theta+\frac{\pi}{3})=
\lambda^{(m)}(\hat{C}_{2x})\hat{C}_{2x}v_P^{-\nu(m)}(\theta)\; , \\
v_P^{\nu(m)}(-\theta)=
\lambda^{(m)}(\hat{\sigma}_{xz})\hat{\sigma}_{xz}v_P^{\nu(m)}(\theta)\; ,
\end{array} \right.
\label{vector}\end{equation}
respectively, in which $\lambda^{(m)}(\hat{o})$ is the phase
factor taking value 1 or $-1$ corresponding to mode $m$ such as
$\vec{e}_r$ AC, $\vec{e}_\phi$ OP and $\vec{e}_z$ OP and referring
to the symmetry operation $\hat{o}$ such as $\hat{C}_{2x}$ and
$\hat{\sigma}_{xz}$. In particular, we have
$$\begin{array}{c@{\:=\:}r@{\:=\:}r}
\lambda^{(\vec{e}_r \mbox{\scriptsize AC})}(\hat{C}_{2x})      &
\lambda^{(\vec{e}_r \mbox{\scriptsize AC})}(\hat{\sigma}_{xz}) & 1\;, \\
\lambda^{(\vec{e}_\phi \mbox{\scriptsize OP})}(\hat{C}_{2x})   &
\lambda^{(\vec{e}_\phi \mbox{\scriptsize OP})}(\hat{\sigma}_{xz}) & -1\;, \\
\lambda^{(\vec{e}_z \mbox{\scriptsize OP})}(\hat{C}_{2x})         &
-\lambda^{(\vec{e}_z \mbox{\scriptsize OP})}(\hat{\sigma}_{xz})   & -1\;.
\end{array}  $$
It can be easily verified that the parameterized expressions for
the modes $\vec{e}_r$ AC, $\vec{e}_\phi$ OP and $\vec{e}_z$ OP
listed in Table~\ref{Tab:eigenvector} satisfy exactly
Eq.~(\ref{vector}). Interestingly, the values taken by
$\lambda^{(m)}(\hat{o})$ coincide precisely with the signatures
of the major component of the corresponding modes with respect to
the operations $\hat{C}_{2x}$ and $\hat{\sigma}_{xz}$. The detailed
expressions for polarization vectors are given in the appendix.

It is known that the planar graphene is the $r\rightarrow\infty$
limit of the SWCNT. We notice that six modes in
Table~\ref{Tab:omega} evolve to the two degenerate in-plane
optical modes of the graphene  with approximately the same
frequency limit $1594.0$. While  three modes with
$f_{0}(r)\rightarrow 864.8$  approach the out-of-plane optical
mode of the graphene sheet. Moreover shown in
Table~\ref{Tab:velocity}, the sound velocities of the two zero
modes, i.e. LA and TW modes belonging to $_0A_{0}^{-}$, have
nonzero limits with different values. Therefore, it is expected
that those two modes would approach two acoustic modes of the
graphene sheet. We further notice that the frequency of the
$\vec{e}_r$ AC mode belonging to $_0A_{0}^{+}$
(Table~\ref{Tab:omega}) vanishes with its polarization vector
perpendicular to the limiting sheet. We might interpret that this
mode is a kind of precursor of the flexure mode of the graphitic
sheet. However, at $(\kappa, n)=\pm(\alpha, 1)$, the dispersion of
the TA branch is quadratic in $\kappa$. We stress that the
parameterization for the coefficient $\beta$ cannot be
extrapolated to $r\rightarrow\infty$. This is prohibited by a kind
of symmetry argument that the rod-like tube has two flexure modes
with the cylindrical symmetry while the plate-like graphene sheet
breaks the symmetry so as to have only one flexure mode. There is
no way to cross continuously from the former  to the latter.

The sound velocities, slopes of the dispersions,  of zero modes
can be determined correctly only if we have enough data of the
dispersions with high precision in the low frequency limit around
$(\kappa,n)=(0,0)$. As one of the immediate consequences of our
calculations, besides that for $C_{\mbox{\tiny LA}}$, our
calculations provide a credible value for $C_{\mbox{\tiny TW}}$ by
taking the advantage that our model keeps exactly the
translational and rotational invariance. With the velocities
$C_{\mbox{\tiny LA}}$ and $C_{\mbox{\tiny TW}}$ we can extract the
chirality-dependent expression for the Poisson ratio
$\mu$.\cite{Popov2} Ignoring $\frac{1}{r^{4}}$ and higher order
terms, we get
\begin{eqnarray}
    \mu=0.21+\frac{0.206}{r^{2}}-\frac{0.43}{r^{2}}\cos6\theta \;.
    \label{Poisson}
\end{eqnarray}
We may take the values of $\theta$  as either $\frac{\pi}{6}$
(armchair) or 0  (zigzag) in the above Eq.~(\ref{Poisson}) to
extract the Poisson ratios for the achiral cases:
$$ \begin{array}{c@{, \qquad \mbox{for} \quad}c}
\mu=0.21+\frac{0.636}{r^{2}} & \theta=\frac{\pi}{6} \; ; \\
\mu=0.21-\frac{0.224}{r^{2}} & \theta=0 \; .
\end{array} $$
The corresponding results reported in Ref.~\onlinecite{Popov2}
derived by the perturbation approach can be fitted by our
expression in good agreement.

\section{conclusions}
In conclusion, we presented a detailed expression for the
vibrational potential of the SWCNT. On the one hand, they obey
precisely the rigid rotational and translational symmetries, in
which the bond length differences as well as those of bond angles
are properly considered following the geometry of cylindrically
warped graphene sheet. On the other hand, they exhibit the chiral
symmetry inherited from the planar hexagonal lattice. With this
model we calculated the phonon spectrum of the SWCNT and
parameterized the physical quantities in terms of tube diameter
and chiral angle to make the chiral symmetry to be explicit. Our
calculated results are mainly illustrated and summarized in Fig.~4
and Tables~\ref{Tab:omega}, \ref{Tab:velocity}, and
\ref{Tab:eigenvector}. The physics of relevant quantities such as
flexure modes, some of the immediate consequences like Poisson
ratio, and the symmetry properties of, for example, the
polarization vectors are discussed. Our paper provides a
comprehensive and systematic understanding for the lattice
dynamics of the SWCNT, and forms a kind of sound basis for further
improvements and developments.

\appendix
\section{the expression for physical quantity}
With the symmetry restriction as Eqs~(\ref{C2x}) and
(\ref{sigmaxz}), i.e.
$\lambda^{(m)}(C_{2x})=\lambda^{(m)}(\sigma_{xz})=1$ in
Eq.~(\ref{vector}), the general expression
Eq.~(\ref{Fourierexpansion}) would reduce to the following.

Since the frequency $\omega$ is a scalar, we have
\begin{eqnarray}
  \begin{array}{l}
\omega^{\pm}(\theta)=a_0\pm a_1\cos(3\theta)+a_2\cos(6\theta)\pm a_3\cos(9\theta)+\ldots;\\
\omega^{0}(\theta)=a_0+a_2\cos(6\theta)+\ldots \;.
  \end{array}
\end{eqnarray}
The three components of a vector can be expanded as
\begin{eqnarray}
    v_{r}^{\nu}(\theta)&=&\sum_{n=0}^{\infty}\nu^{[\frac{1-(-1)^{n}}{2}]}a_{n}\cos3n\theta \nonumber\\
    &=&a_{0}+\nu a_{1}\cos3\theta+a_{2}\cos6\theta+\ldots \; ,
    \nonumber\\
    v_{\phi}^{\nu}(\theta)&=&\sum_{n=0}^{\infty}\nu^{[\frac{1+(-1)^{n}}{2}]}b_{n}\sin3n\theta \nonumber\\
    &=&b_{1}\sin3\theta+\nu b_{2}\sin6\theta+\ldots \; ,
    \nonumber\\
    v_{z}^{\nu}(\theta)&=&\sum_{n=0}^{\infty}\nu^{[\frac{1+(-1)^{n}}{2}]}a_{n}\cos3n\theta \nonumber\\
    &=&\nu a_{0}+a_{1}\cos3\theta+\nu a_{2}\cos6\theta+\ldots \; .
    \label{vector2}
\end{eqnarray}

Moreover, for polarization vectors, when
$\lambda^{m}(\sigma_{xz})=-1$, and $\lambda^{m}(C_{2x})=-1$, we
have
\begin{eqnarray}
    v_{r}^{\nu}(\theta)&=&\sum_{n=0}^{\infty}\nu^{[\frac{1+(-1)^{n}}{2}]}b_{n}\sin3n\theta \nonumber\\
    &=&b_{1}\sin3\theta+\nu b_{2}\sin6\theta+\ldots \; ,
    \nonumber\\
    v_{\phi}^{\nu}(\theta)&=&\sum_{n=0}^{\infty}\nu^{[\frac{1-(-1)^{n}}{2}]}a_{n}\cos3n\theta \nonumber\\
    &=&a_{0}+\nu a_{1}\cos3\theta+a_{2}\cos6\theta+\ldots \; ,
    \nonumber\\
    v_{z}^{\nu}(\theta)&=&\sum_{n=0}^{\infty}\nu^{[\frac{1-(-1)^{n}}{2}]}b_{n}\sin3n\theta \nonumber\\
    &=&\nu b_{1}\sin3\theta+b_{2}\sin6\theta+\ldots \; .
    \label{vector3}
\end{eqnarray}

When $\lambda^{m}(\sigma_{xz})=1$, and $\lambda^{m}(C_{2x})=-1$,
we have
\begin{eqnarray}
    v_{r}^{\nu}(\theta)&=&\sum_{n=0}^{\infty}\nu^{[\frac{1+(-1)^{n}}{2}]}a_{n}\cos3n\theta \nonumber\\
    &=&\nu a_{0}+a_{1}\cos3\theta+\nu a_{2}\cos6\theta+\ldots \; ,
    \nonumber\\
    v_{\phi}^{\nu}(\theta)&=&\sum_{n=0}^{\infty}\nu^{[\frac{1-(-1)^{n}}{2}]}b_{n}\sin3n\theta \nonumber\\
    &=&\nu b_{1}\sin3\theta+b_{2}\sin6\theta+\ldots \; ,
    \nonumber\\
    v_{z}^{\nu}(\theta)&=&\sum_{n=0}^{\infty}\nu^{[\frac{1-(-1)^{n}}{2}]}a_{n}\cos3n\theta \nonumber\\
    &=&a_{0}+\nu a_{1}\cos3\theta+a_{2}\cos6\theta+\ldots \; .
    \label{vector4}
\end{eqnarray}

When $\lambda^{m}(\sigma_{xz})=-1$, and $\lambda^{m}(C_{2x})=1$,
we have
\begin{eqnarray}
    v_{r}^{\nu}(\theta)&=&\sum_{n=0}^{\infty}\nu^{[\frac{1-(-1)^{n}}{2}]}b_{n}\sin3n\theta \nonumber\\
    &=&\nu b_{1}\sin3\theta+b_{2}\sin6\theta+\ldots \; ,
    \nonumber\\
    v_{\phi}^{\nu}(\theta)&=&\sum_{n=0}^{\infty}\nu^{[\frac{1+(-1)^{n}}{2}]}a_{n}\cos3n\theta \nonumber\\
    &=&\nu a_{0}+a_{1}\cos3\theta+\nu a_{2}\cos6\theta+\ldots \; ,
    \nonumber\\
    v_{z}^{\nu}(\theta)&=&\sum_{n=0}^{\infty}\nu^{[\frac{1+(-1)^{n}}{2}]}b_{n}\sin3n\theta \nonumber\\
    &=&b_{1}\sin3\theta+\nu b_{2}\sin6\theta+\ldots \; .
    \label{vector5}
\end{eqnarray}

\end{document}